\documentclass[prd,aps,showpacs,superscriptaddress,nofootinbib,preprint]{revtex4-1}
\usepackage{amssymb}
\usepackage{graphicx}% Include figure files
\def\slashchar#1{\setbox0=\hbox{$#1$}
   \dimen0=\wd0 \setbox1=\hbox{/} \dimen1=\wd1
   \ifdim\dimen0>\dimen1 \rlap{\hbox to \dimen0{\hfil/\hfil}} #1
   \else  \rlap{\hbox to \dimen1{\hfil$#1$\hfil}} / \fi}

\begin{document}
\title{$\nu(\bar\nu)$-$^{208}$Pb deep inelastic scattering}
\author{H. \surname{Haider}}
\affiliation{Department of Physics, Aligarh Muslim University, Aligarh-202 002, India}
\author{I. \surname{Ruiz Simo}}
\affiliation{Departamento de F\'\i sica Te\'orica and IFIC, Centro Mixto
Universidad de Valencia-CSIC, Institutos de Investigaci\'on de
Paterna, E-46071 Valencia, Spain}
\affiliation{Departamento de F\'isica At\'omica Molecular y Nuclear, Universidad de Granada, E-18071 Granada, Spain}
\author{M. Sajjad \surname{Athar}}
\email{sajathar@gmail.com}
\affiliation{Department of Physics, Aligarh Muslim University, Aligarh-202 002, India}
\begin{abstract}
Nuclear-medium effects in the weak structure functions $F_2(x,Q^2)$ and $F_3(x,Q^2)$ in the charged current neutrino and antineutrino induced deep inelastic reactions in $^{208}$Pb have been studied. 
The calculations have been performed in a theoretical model using relativistic nuclear spectral functions which incorporate Fermi motion, binding and nucleon correlations.
 We also consider the pion and rho meson cloud contributions calculated from a microscopic model for meson-nucleus self-energies. Using these structure functions, the results for the differential cross section 
have been obtained and compared with the CERN Hybrid Oscillation Research apparatUS (CHORUS) data. The results for the
ratios $\frac{2F_{i}^{Pb}}{208F_i^D}$, $\frac{4F_{i}^{Pb}}{208F_i^{He}}$, $\frac{12F_{i}^{Pb}}{208F_i^C}$, $\frac{16F_{i}^{Pb}}{208F_i^O}$, and $\frac{56F_{i}^{Pb}}{208F_i^{Fe}}$ (i=2,3) 
have also been obtained and a few have been compared with some of the phenomenological fits.  
\end{abstract}
\maketitle

\section{Introduction}
In recent years the need for a better understanding of nucleon dynamics in the nuclear medium has been emphasized upon in the weak interaction induced processes in order to precisely estimate the lepton event rates in the 
Monte Carlo generators which are being used in the analysis of present generation of neutrino oscillation experiments~\cite{nufact,nuint}. Most of these experiments are using neutrino 
and antineutrino beams in the few GeV energy region and the targets are nuclei like 
$^{12}$C, $^{56}$Fe, $^{208}$Pb, and so on. In the few GeV energy region the contribution to the cross section comes from the quasielastic, inelastic as well as the deep inelastic processes. In the deep inelastic region,  
both experimentally as well as theoretically, limited efforts 
have been made to understand the medium effects in the structure functions.
Substantial number of experiments have been performed using charged lepton beams to study nuclear modifications of the electromagnetic structure function $F_2^{EM}(x,Q^2)$ with several nuclear targets while few experiments have been performed using neutrino
 and antineutrino beams 
 on some nuclear targets like carbon, neon, iron and lead~\cite{nomad}-\cite{Tzanov}. In the case of experimental measurements performed for $F_{2,3}^{Weak}(x,Q^2)$ the error bars are large and need better precision. In general, nuclear modifications in the weak structure functions $F_{2,3}^{Weak}(x,Q^2)$ may be different from the nuclear modification in 
$F_2^{EM}(x,Q^2)$. Particularly neutrino data are important in the determination of valence quark distributions in the nucleon. This is due to the fact that the parity-violating
($F_3^{Weak}$) structure function directly probes into the valence distributions. Precise determination of parton distribution functions (PDFs) is also necessary for the new physics at the colliders. But for the determination 
of PDFs, nuclear medium effects should be properly accounted for.
Nuclear-medium effects may be responsible for the anomaly observed in the NuTeV experiment for the weak mixing angle sin$^2\theta_W$~\cite{Zeller}. 
NuTeV Collaboration~\cite{Tzanov} has performed measurements of neutrino and antineutrino induced deep inelastic scattering cross sections $\frac{d^2\sigma}{dx dy}$ using iron target. The differential cross sections
are then used to extract weak structure functions $F_2^{Weak}(x,Q^2)$ and $F_3^{Weak}(x,Q^2)$. CHORUS Collaboration~\cite{chorus1, chorus2} has performed high-statistics measurement of the
differential (anti)neutrino cross sections using mainly lead target at various neutrino and antineutrino energies as a function of Bjorken scaling variables x and y. 
Analysis is being made at the Neutrino Oscillation MAgnetic Detector (NOMAD)~\cite{nomad} for the weak structure functions and cross section measurements with carbon target using neutrino beam. 
 The Neutrino Scattering On Glass (NuSOnG) experiment~\cite{nusong}
has been proposed at Fermilab to study the structure functions
in the deep inelastic region using neutrino scattering on carbon target. Main INjector Experiment $\nu$-A (MINER$\nu$A)~\cite{minerva} is taking data using neutrinos
from the Neutrinos at the Main Injector (NuMI) facility, measuring neutrino cross sections in the energy region of 1-20 GeV using various nuclear targets and the aim is to study the strong dynamics of the nucleon and nucleus 
that affect these interactions as well as the parton distribution functions. 
 The Oscillation Project with Emulsion-tRacking Apparatus (OPERA) experiment~\cite{opera} is a long baseline experiment using lead emulsion target and the main purpose of the experiment
is the observation of $\nu_\mu$ to $\nu_\tau$ oscillations in the direct appearance mode.  
The charged current deep inelastic scattering process is found to be dominant in this experiment with a fraction exceeding 90$\%$. Therefore, several experimental activities are going on to study neutrino as well as antineutrino event rates in the 
deep inelastic region.

On the theoretical side, for the weak interaction induced processes in the deep inelastic region, the dynamical origin of the nuclear medium effects has been studied by a few
 authors~\cite{Petti, Sajjad, Sajjad2}. In some other theoretical analysis, nuclear medium effects have been phenomenologically described in terms of a few parameters which are determined by 
fitting the experimental data of charged leptons and (anti)neutrino deep inelastic scattering from various nuclear targets~\cite{Hirai, hirai, eskola, Bodek1, Eskola, Kovarik, Kovarik1, Schienbein1}. 
The various phenomenological studies differ in the choice of the data sets(Lepton-Nucleus+Drell Yan data,  Lepton-Nucleus+Drell Yan+$\nu(\bar\nu)$-Nucleus data, and $\nu(\bar\nu)$-Nucleus data), experimental cuts in their analysis, parametrization of the parton distributions, etc.

In the present work, we study nuclear medium  effects on the structure functions $F_2$($x$,$Q^2$) and $F_3$($x$,$Q^2$) in lead, treating it as a nonisoscalar nuclear target. 
 This study has been performed using a relativistic nuclear spectral function~\cite{FernandezdeCordoba:1991wf} to describe the momentum distribution of nucleons in the nucleus within a field-theoretical approach where nucleon 
propagators are written in terms of this spectral function, and nuclear many-body theory is used to calculate it for an interacting Fermi sea in nuclear matter.
A local-density approximation is then applied to translate these results to finite nuclei~\cite{Marco, Sajjad, Sajjad1}. 

We have assumed the Callan-Gross relationship for nuclear structure functions ${F_1}^A(x,Q^2)$ and
 ${F_2}^A(x,Q^2)$. The contributions of the pion and rho meson clouds are taken into account in a many-body field-theoretical approach which is  
based on Refs.~\cite{Marco,GarciaRecio:1994cn}. We have taken into account target mass correction (TMC) following Ref.~\cite{schienbein} which has a significant effect at low $Q^2$, moderate, and high 
Bjorken $x$. To take into account the shadowing effect, which is important at low $Q^2$ and low x, and modulates the contribution of pion and rho cloud contributions, we have followed the works of 
Kulagin and Petti~\cite{Kulagin, Petti}. Earlier we have applied the present formalism to study nuclear effects in the electromagnetic structure function $F_{2}^{EM}(x, Q^2)$ in nuclei 
in the charged lepton-nucleus deep inelastic scattering~\cite{Sajjad1} as well as to the study of weak structure functions $F_i(x,Q^2)$ (i=2,3) in carbon and iron nuclei~\cite{Sajjad2} 
using (anti)neutrino-nucleus scattering. In the case of electromagnetic interaction we have found that
the results of our calculations are in agreement with recent results as obtained from the Thomas Jefferson National Accelerator Facility (JLab)~\cite{Seely} and also with some 
of the earlier experiments performed using heavier nuclear targets. While in the case of neutrino and antineutrino induced reactions, the results in carbon and iron were compared with the experimental 
results of NuTeV~\cite{Tzanov} 
and the CERN Dortmund Heidelberg Saclay Warsaw (CDHSW)~\cite{Berge} data and found to be in agreement. In this paper, we have studied nuclear-medium effects in the neutrino and antineutrino induced deep 
inelastic scattering on lead, and obtained the results for the weak structure functions
 $F_2^A$($x$,$Q^2$) and $F_3^A$($x$,$Q^2$). Using our results for the structure functions in helium, carbon, oxygen and iron~\cite{Sajjad2}, we have obtained the results for 
$\frac{2F_{i}^{Pb}}{208F_i^D}$, $\frac{4F_{i}^{Pb}}{208F_i^{He}}$, $\frac{12F_{i}^{Pb}}{208F_i^C}$, $\frac{16F_{i}^{Pb}}{208F_i^O}$ and $\frac{56F_{i}^{Pb}}{208F_i^{Fe}}$ (i=2,3). 
These results may be quite useful in the analysis
 of the MINER$\nu$A~\cite{minerva} experiment. Finally the structure functions in lead are used to obtain the differential 
scattering cross section and the results are compared with the experimental results from CHORUS~\cite{chorus2}.
 
The plan of the paper is as follows. In Sect.~\ref{Formalism} we present in brief the formalism for $\nu(\bar{\nu})$-nucleus scattering,
in Sect.~\ref{Sec:Results} we present and discuss the results of our calculations and compare them with the available experimental results. In Sect.~\ref{Sec:Conclusion} we conclude our findings.
\section{Formalism}\label{Formalism}
The differential cross section for charged current neutrino (antineutrino) interaction with a nucleus is written as:
\begin{equation}
\frac{d^2\sigma^{\nu(\bar{\nu})A}_{CC}}{dE'\;d\Omega'}=\frac{G^2_F}{(2\pi)^2} \; \frac{\left|\vec{k}'\right|}{\left|\vec{k}\right|} \; \left(\frac{m^2_W}{q^2-m^2_W}\right)^2 \;
 L^{\alpha\beta}_{\nu,\bar{\nu}} \; W^{\nu(\bar{\nu})A}_{\alpha\beta}
\end{equation}
where $E'$ and $\Omega'$ are the energy and scattering angles ($\theta'$,$\phi'$) of the outgoing lepton; $G_F$ is the 
Fermi constant of the weak interaction; $\vec{k}'$ and $\vec{k}$ are the momenta of the outgoing and incoming leptons, respectively; $q^2=(k\;-\;k^\prime)^2$ is the four momentum transfer square, 
$m_W$ is the mass of the boson $W^\pm$. $L^{\alpha\beta}_{\nu,\bar{\nu}}$ and $W^{\nu(\bar{\nu})A}_{\alpha\beta}$ are the leptonic and the hadronic tensors given by:
\begin{eqnarray}
L^{\alpha\beta}_{\nu,\bar{\nu}}(k,k')&=&k^\alpha k'^\beta+k'^\alpha k^\beta-g^{\alpha\beta}(k\cdot k')\mp i\;\epsilon^{\alpha\beta\mu\nu}k_\mu k'_\nu \label{eq:leptonic_tensor}\\
W^{\nu(\bar{\nu})A}_{\alpha\beta}&=&\left(\frac{q_\alpha q_\beta}{q^2}-g_{\alpha\beta}\right)W^{\nu(\bar{\nu})A}_1(P_A,q)\nonumber\\
&+&\frac{1}{M^2_A}\left(P_{A\alpha}-\frac{P_A\cdot q}{q^2}q_\alpha\right)\left(P_{A\beta}-\frac{P_A\cdot q}{q^2}q_\beta\right)W^{\nu(\bar{\nu})A}_2(P_A,q)\nonumber\\
&-&\frac{i}{2M^2_A}\;\epsilon_{\alpha\beta\rho\sigma}\; P^\rho_Aq^\sigma W^{\nu(\bar{\nu})A}_3(P_A,q)+\frac{1}{M^2_A}q_\alpha q_\beta W^{\nu(\bar{\nu})A}_4(P_A,q)\nonumber\\
&+&\frac{1}{M^2_A}\left(P_{A\alpha}q_\beta+q_\alpha P_{A\beta}\right)W^{\nu(\bar{\nu})A}_5(P_A,q)\label{eq:nuclear_hadronic_tensor}
\end{eqnarray}
where $P_A$ is the momentum of the nucleus A and $W^A_{i}(x,Q^2)$ (i=1-5) nuclear hadronic structure functions. $M_A$ is the mass of the nucleus. In the antisymmetric term of the leptonic tensor, - sign is for the neutrino 
and + sign for the antineutrino induced reaction. $W_4$ and $W_5$ are generally omitted in the cross section expression since they are suppressed by a factor which vanishes 
in the limit $m_l \to 0$, being $m_l$ the mass of the outgoing charged lepton. Finally $W^A_i(x,Q^2)$ (i=1-3) are redefined in terms of the dimensionless structure functions $F^A_{i}(x,Q^2)$ through 
\begin{footnotesize}
\begin{eqnarray}\label{relation}
M_A W_1^A(\nu, Q^2)=F_1^A(x, Q^2),~~\nu W_2^A(\nu, Q^2)=F_2^A(x, Q^2),~~\nu W_3^A(\nu, Q^2)=F_3^A(x, Q^2) 
\end{eqnarray}
\end{footnotesize}
We have also assumed the Callan-Gross relationship for nuclear structure functions ${F_1}^A(x)$ and
 ${F_2}^A(x)$, therefore, we are left with only two independent structure functions viz. ${F_2}^A(x)$ and
 ${F_3}^A(x)$.

For the numerical calculations, parton distribution
 functions for the nucleons have been taken from the parametrization of the Coordinated Theoretical-Experimental Project on QCD
(CTEQ) Collaboration (CTEQ6.6)~\cite{cteq}. The Next-to-Leading-Order (NLO) evolution of the deep inelastic structure functions has been taken from the works of 
Moch et al.~~\cite{Vermaseren,Moch}. 

In the local-density approximation the nuclear hadronic tensor $W^{\nu(\bar{\nu})A}_{\alpha\beta}$ can be written as a convolution of the nucleonic hadronic tensor
 with the hole spectral function. For symmetric nuclear matter, this would be:

\begin{equation}\label{iso}
W^{\nu(\bar{\nu})A}_{\alpha\beta}=4\int d^3r\int\frac{d^3p}{(2\pi)^3}\frac{M}{E(\mathbf{p})}\int^{\mu}_{-\infty}dp^0\; S_{hole}(p^0,\mathbf{p},k_F(\vec{r}))~ W^{\nu(\bar{\nu})N}_{\alpha\beta}
\label{eq:convolution_hadronic_tensor_symm_nuclear_matter}
\end{equation}
where $k_F(\vec{r})$ is the Fermi momentum for symmetric nuclear matter, depending on the density of nucleons in the nucleus, i.e. $k_F(\vec{r})=\left(\frac{3\pi^2}{2}\rho(\vec{r})\right)^{1/3}$.
 
For a non-symmetric nucleus like $^{208}$Pb,
one considers separate distributions of Fermi seas for protons and neutrons and the above expression modifies to:
\begin{eqnarray}\label{noniso}
W^{\nu(\bar{\nu})A}_{\alpha\beta}&=&2\int d^3r\int\frac{d^3p}{(2\pi)^3}\frac{M}{E(\mathbf{p})}\int^{\mu_p}_{-\infty}dp^0\; S^{proton}_{h}(p^0,\mathbf{p},k_{F,p}) ~ W^{\nu(\bar{\nu})p}_{\alpha\beta}\nonumber\\
&+&2\int d^3r\int\frac{d^3p}{(2\pi)^3}\frac{M}{E(\mathbf{p})}\int^{\mu_n}_{-\infty}dp^0\; S^{neutron}_{h}(p^0,\mathbf{p},k_{F,n})~ W^{\nu(\bar{\nu})n}_{\alpha\beta}
\end{eqnarray}
where the factor $2$ in front of the integral accounts for the two degrees of freedom of the spin of the nucleons.
 There are two different spectral functions, each one of them normalized to the number of protons or neutrons in the nuclear target and are functions of Fermi momentum of protons
 and neutrons respectively which are given by $k_{F,p}=(3\pi^2\rho_p)^{1/3}$ and $k_{F,n}=(3\pi^2\rho_n)^{1/3}$.  For the proton and neutron densities in lead, we have used 
two-parameter Fermi density distribution given by
\begin{equation} \label{density2}
\rho(r)=\frac{\rho_0}{1 + \exp(\frac{r-c1}{c2})},
\end{equation}
 where the parameters c1=6.624 fm, c2=0.549 fm for the protons and c1=6.89 fm and c2=0.549 fm for the neutrons have been used~\cite{Density}. We have taken 3-parameter Fermi density for Helium, 
harmonic oscillator densities for carbon and oxygen and 2-parameter Fermi density for iron nuclei and the density parameters are taken from Refs.~\cite{Density, Vries}.
\begin{figure}[t]
\includegraphics[width=12cm,height=6cm]{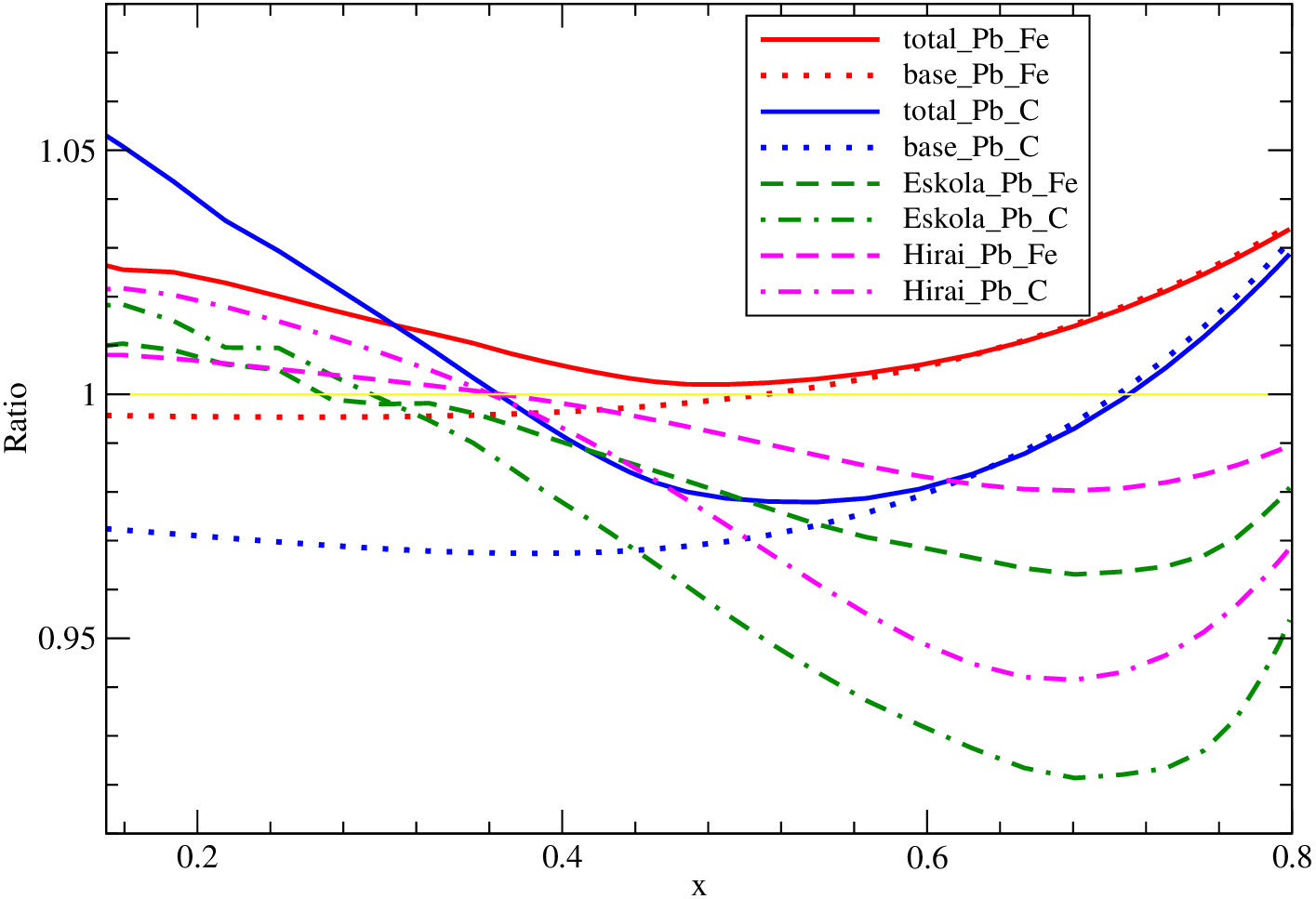}
\caption{(Color online) Ratio R(x,$Q^2$)=$\frac{12F_{2}^{Pb}}{208F_2^C}$ and R(x,$Q^2$)=$\frac{56F_{2}^{Pb}}{208F_2^{Fe}}$ using our base equation with TMC (dotted line) and the full model (solid line) at LO for 
$Q^2=5$ GeV$^2$. The results from 
Hirai et al.~\cite{hirai} and Eskola et al.~\cite{Eskola} have also been shown.\\}
\label{f2_ratio_lo.eps}
\end{figure}
%\vspace{3mm}
\begin{figure}[h]
\includegraphics[width=12cm,height=6cm]{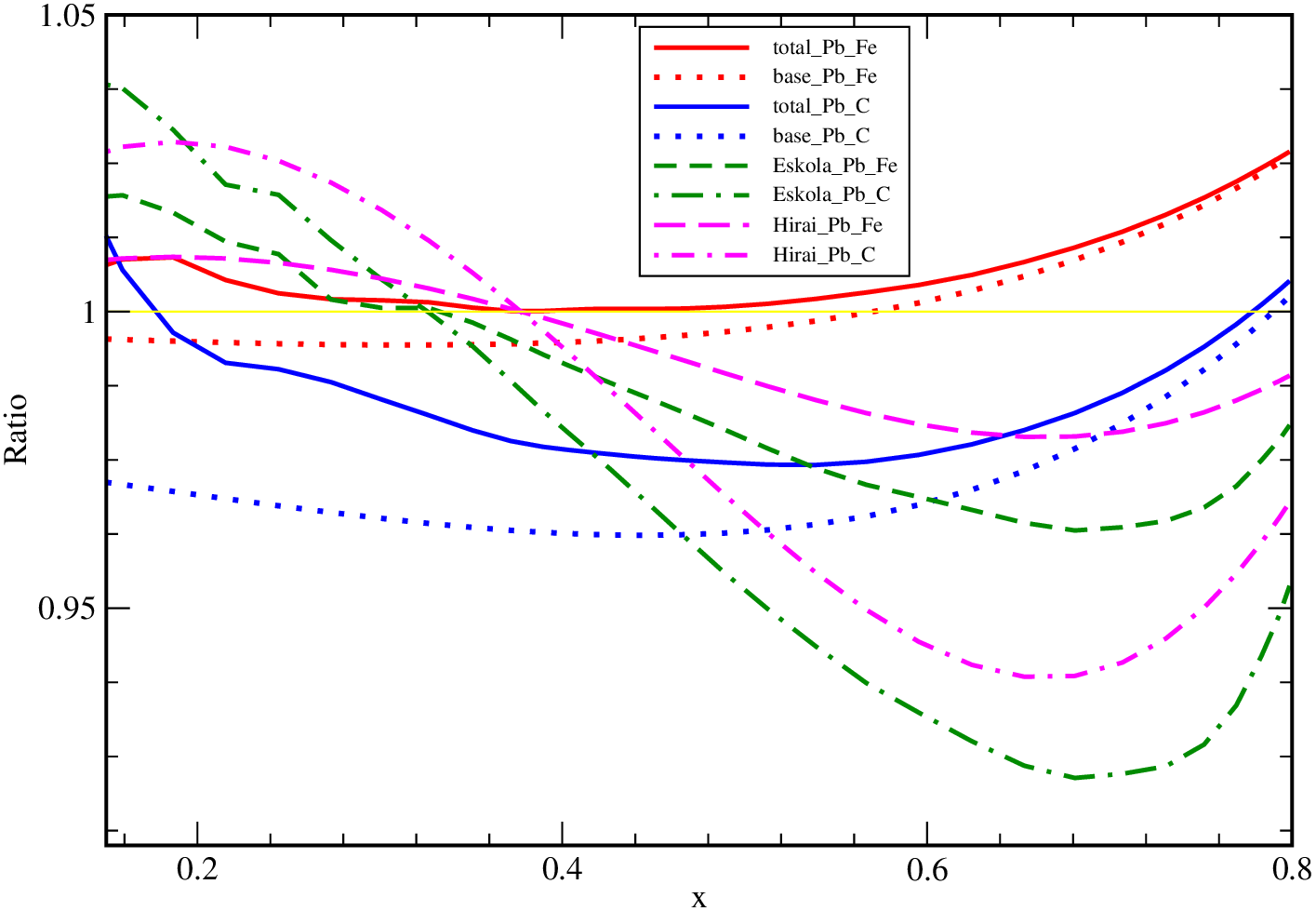}
\caption{(Color online) Ratio R(x,$Q^2$)=$\frac{12F_{3}^{Pb}}{208F_3^C}$ and R(x,$Q^2$)=$\frac{56F_{3}^{Pb}}{208F_3^{Fe}}$. Lines have the same meaning as in Fig.\ref{f2_ratio_lo.eps} \\}
\label{f3_ratio_lo.eps}
\end{figure}
\begin{figure}[h]
\includegraphics[width=12cm,height=6cm]{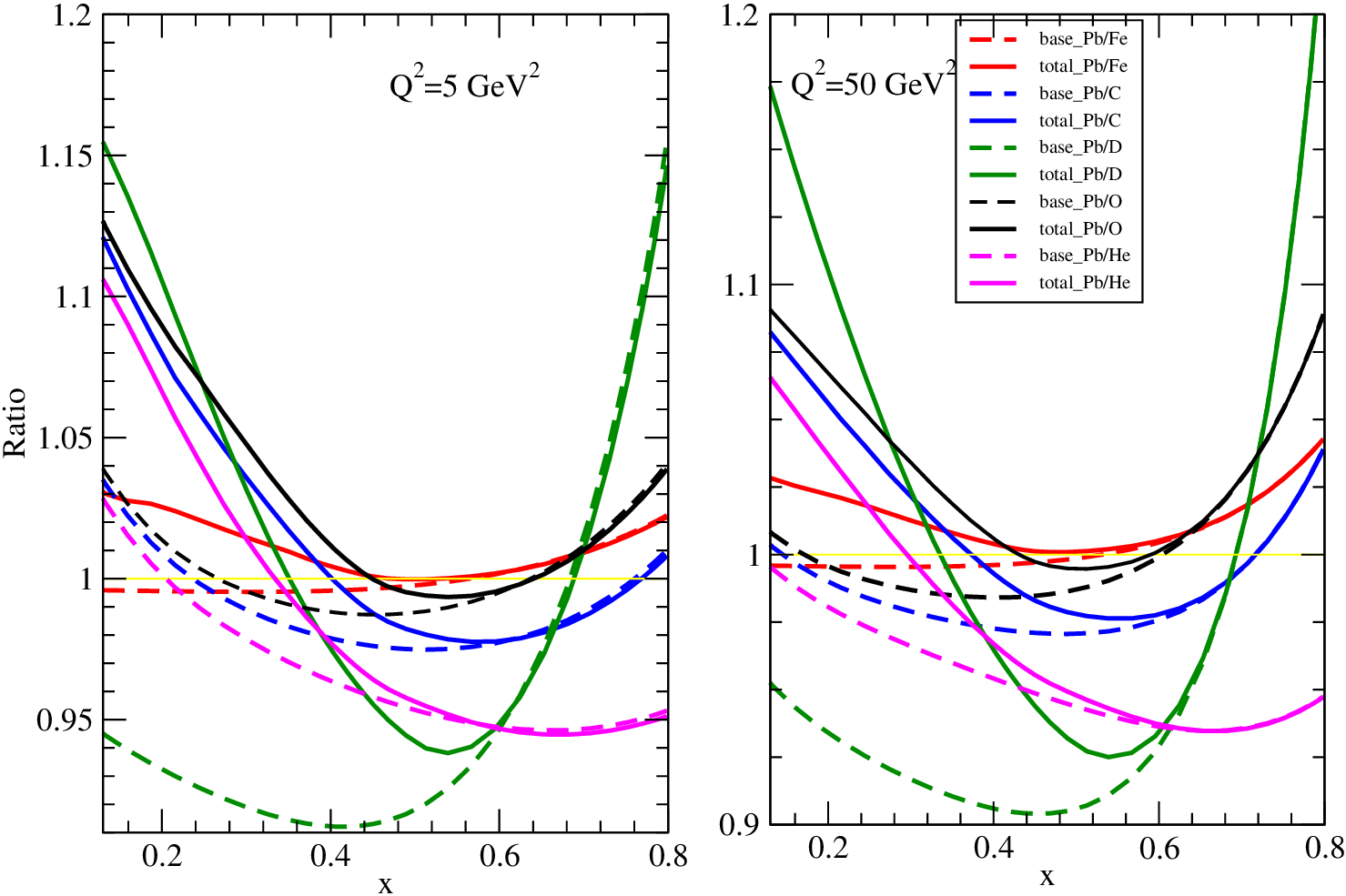}
\caption{(Color online) Ratio R(x,$Q^2$)=$\frac{2F_{2}^{Pb}}{208F_2^D}$ with full calculation for $Q^2=5$ GeV$^2$ and $Q^2=50$ GeV$^2$ using our base equation with TMC (dashed line) and the full model (solid line) at NLO. 
Here the results for the ratios R(x,$Q^2$)=
$\frac{4F_{2}^{Pb}}{208F_2^{He}}$, $\frac{12F_{2}^{Pb}}{208F_2^C}$, $\frac{16F_{2}^{Pb}}{208F_2^O}$ and $\frac{56F_{2}^{Pb}}{208F_2^{Fe}}$ have also been shown for the two cases.\\}
\label{f2_ratio_nlo.eps}
\end{figure}
%\vspace{3mm}
\begin{figure}[h]
\includegraphics[width=12cm,height=6cm]{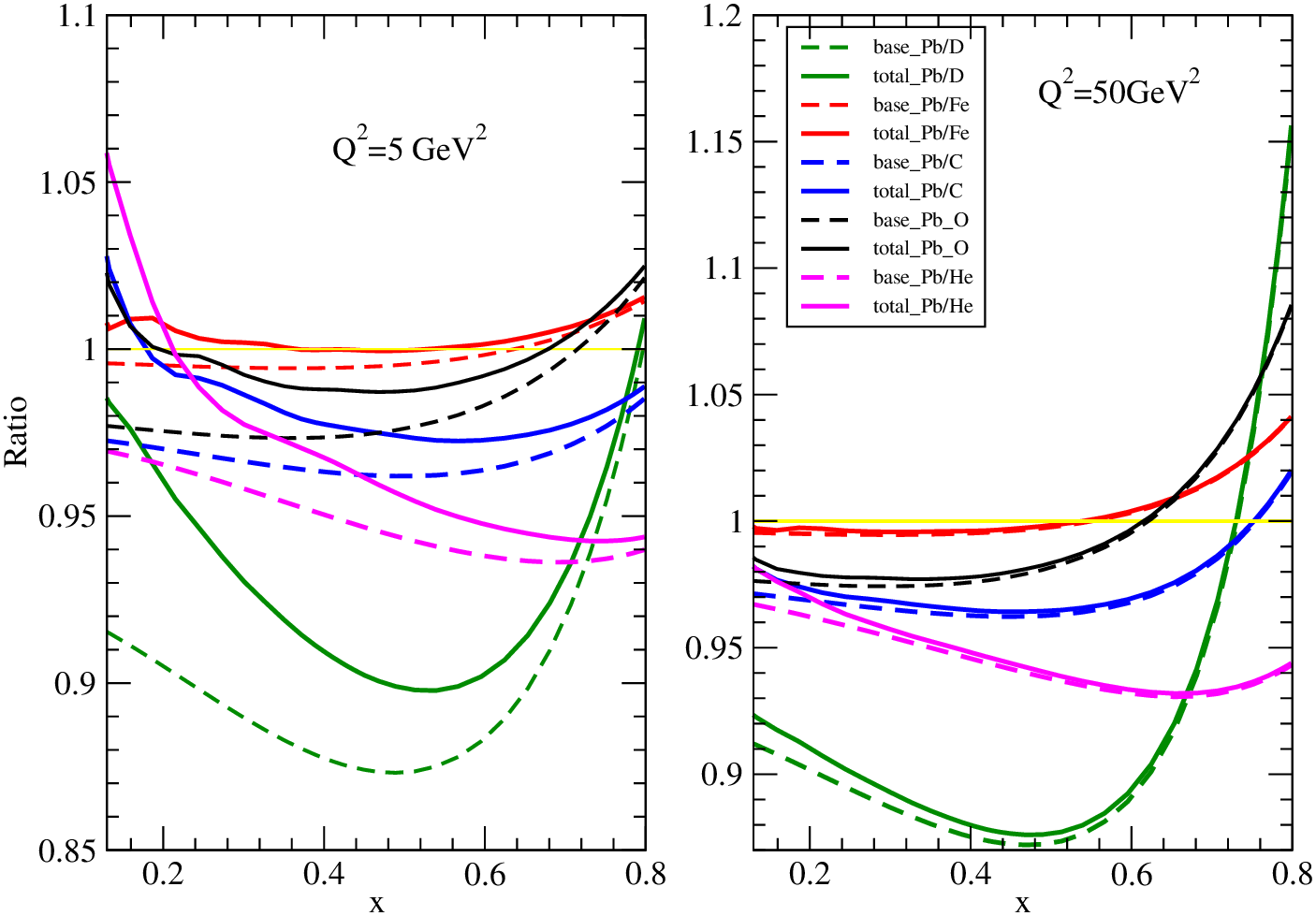}
\caption{(Color online) Ratio R(x,$Q^2$)=$\frac{2F_{3}^{Pb}}{208F_3^D}$ with full calculation for $Q^2=5$ GeV$^2$ and $Q^2=50$ GeV$^2$ using our base equation with TMC (dashed line) and the full model(solid line) at NLO. 
Here the results for the ratios R(x,$Q^2$)=
$\frac{4F_{3}^{Pb}}{208F_3^{He}}$, $\frac{12F_{3}^{Pb}}{208F_3^C}$, $\frac{16F_{3}^{Pb}}{208F_3^O}$ and $\frac{56F_{3}^{Pb}}{208F_3^{Fe}}$ have also been shown for the two cases.\\}
\label{f3_ratio_nlo.eps}
\end{figure}
The invariant quantities for the deep inelastic scattering (DIS) of neutrinos with nuclei are:
\begin{eqnarray}
Q^2=-q^2;\ \ x_A=\frac{Q^2}{2P_A\cdot q};\ \ \nu_A=\frac{P_A\cdot q}{M_A};\ \ y_A=\frac{P_A\cdot q}{P_A\cdot k}
\label{eq:invariant_quantities}
\end{eqnarray}
where $x_A$ is the natural Bjorken variable in the nucleus and $x_A\in\left[0,1\right]$; $y_A$ is the inelasticity. These two variables are related to the nucleonic ones via:
\begin{eqnarray}
x_A&=&\frac{x}{A};\ \ \ \text{where}\ \ \ x=\frac{Q^2}{2Mq^0}\label{eq:xA_x/A}\;\; \text{and} \;\;
y_A=\frac{q^0}{E_\nu}=y\label{eq:yA_y}
\end{eqnarray}
where $x$ is the Bjorken variable for neutrino-nucleon interaction expressed in the nucleon rest frame. We can see that $x\in\left[0,A\right]$, though for $x>1$ the nuclear structure functions are negligible. The variable $y_A$ varies between the following limits:
\begin{equation}
0\leq y_A\leq\frac{1}{1+\frac{M_Ax_A}{2E_\nu}}\approx\frac{1}{1+\frac{Mx}{2E_\nu}}
\label{eq:limits_of_yA}
\end{equation}
therefore, for sufficient high neutrino energy we have $0\leq y_A\leq1$.\\
\begin{figure}[h]
\includegraphics[width=12cm,height=12cm]{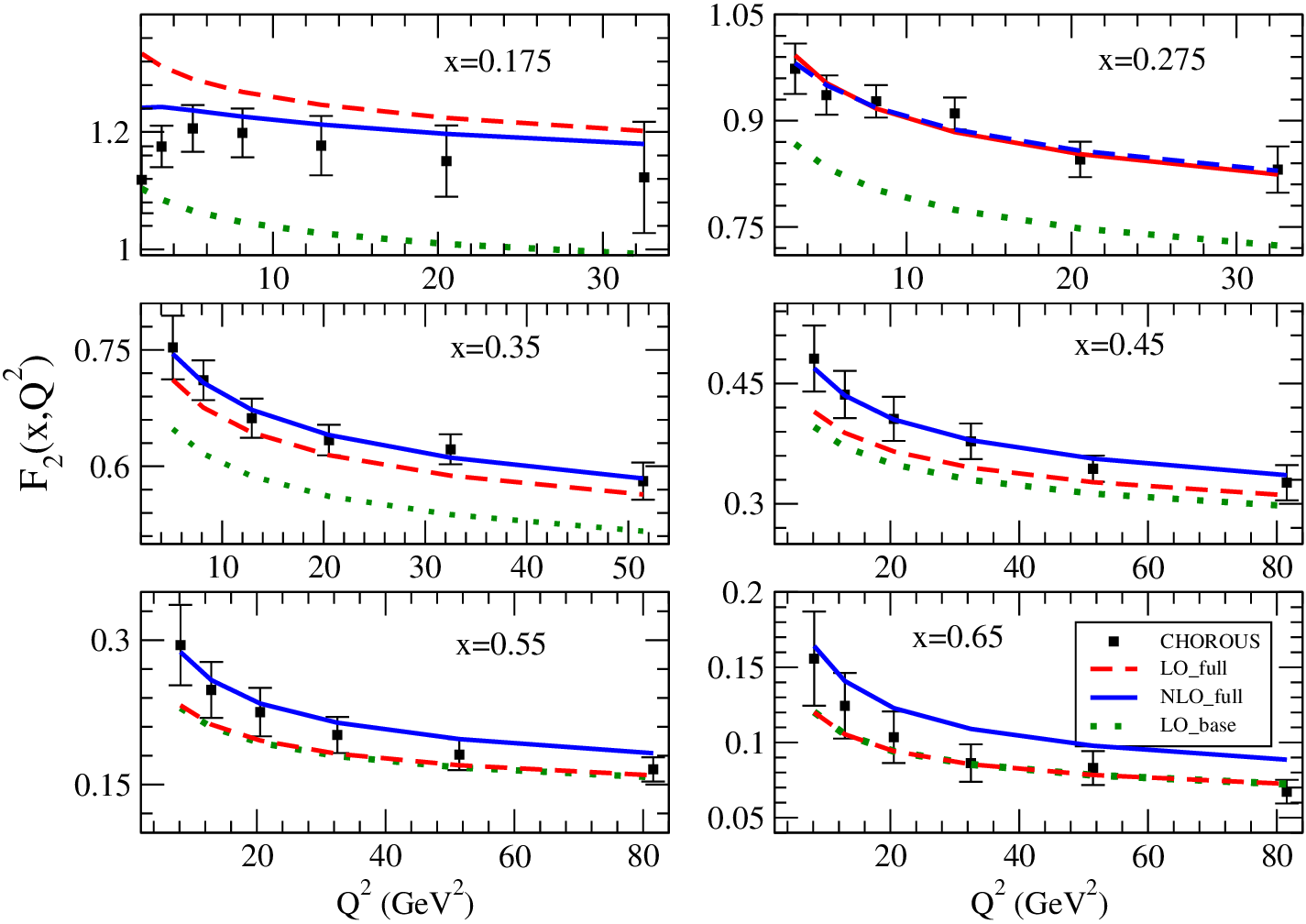}
\caption{(Color online) $F_2(x,Q^2)$ vs $Q^2$ in $^{208}$Pb calculated using our base equation with TMC at LO, the full model at LO and NLO.}
\label{f2.eps}
\end{figure}
%\vspace{3mm}
\begin{figure}[h]
\includegraphics[width=12cm,height=12cm]{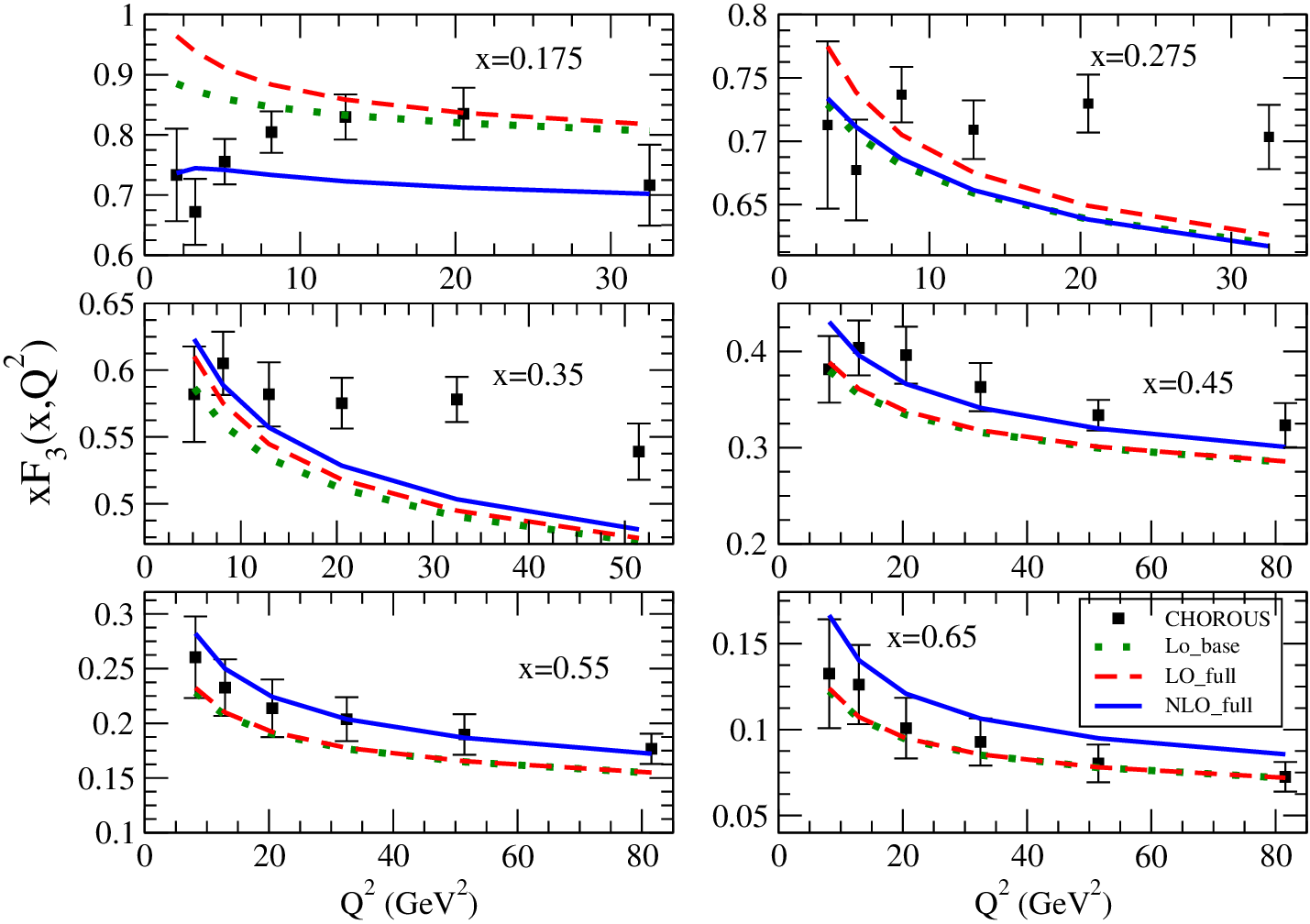}
\caption{(Color online) $xF_3(x,Q^2)$ vs $Q^2$ in $^{208}$Pb calculated using our base equation with TMC at LO, the full model at LO and NLO.}
\label{f3.eps}
\end{figure}

If we express the differential cross section with respect to these variables ($x_A,y_A$), we obtain the following expression in terms of the nuclear structure functions:
\begin{eqnarray}\label{Diff_CS}
\frac{d^2\sigma^{\nu(\bar{\nu})A}_{CC}}{dx_A\;dy_A}&=&\frac{G^2_FM_AE_\nu}{\pi}\left(\frac{m^2_W}{Q^2+m^2_W}\right)^2\Bigg[y^2_Ax_AF^{\nu(\bar{\nu})A}_{1}\nonumber\\
&+&\left(1-y_A-\frac{M_Ax_Ay_A}{2E_\nu}\right)F^{\nu(\bar{\nu})A}_{2}\pm x_Ay_A\left(1-\frac{y_A}{2}\right)F^{\nu(\bar{\nu})A}_{3}\Bigg]\label{eq:CC_cross_section_xA_yA}
\end{eqnarray}
The expressions for $F^A_2(x)$ and $F^A_3(x)$ are obtained as:
\begin{eqnarray}\label{f2Anuclei}
F^A_2(x_A,Q^2)&=& 2\int d^3r\int\frac{d^3p}{(2\pi)^3}\frac{M}{E(\mathbf{p})}\left[ \int^{\mu_p}_{-\infty}dp^0\; S^{proton}_{h}(p^0,\mathbf{p},k_{F,p}) F_2^{proton}(x_N,Q^2) \right. \nonumber   \\ 
 &+& \left. \int^{\mu_n}_{-\infty}dp^0\; S^{neutron}_{h}(p^0,\mathbf{p},k_{F,n}) F_2^{neutron}(x_N,Q^2) \right] \frac{x}{x_N} \left(1+\frac{2x_N p_x^2}{M\nu_N}\right)  
\end{eqnarray}
\begin{eqnarray}\label{f3Anuclei}
F_3^A(x_A,Q^2)&=& 2\int d^3r\int\frac{d^3p}{(2\pi)^3}\frac{M}{E(\mathbf{p})}\left[\int^{\mu_p}_{-\infty}dp^0\; S^{proton}_{h}(p^0,\mathbf{p},k_{F,p}) F_3^{proton}(x_N,Q^2) \right. \nonumber \\
 &+& \left. \int^{\mu_n}_{-\infty}dp^0\; S^{neutron}_{h}(p^0,\mathbf{p},k_{F,n}) F_3^{neutron}(x_N,Q^2) \right] \frac{p^0\gamma-p_z}{(p^0-p_z\gamma)\gamma} 
\end{eqnarray}
where 
\begin{equation}	\label{gamma}
\gamma=\frac{q_z}{q^0}=
\left(1+\frac{4M^2x^2}{Q^2}\right)^{1/2}\, ,  x_N=\frac{Q^2}{2(p^0q^0-p_zq_z)}
\end{equation}

\section{Results and Discussion}\label{Sec:Results}
\begin{figure}
\includegraphics[scale=0.6]{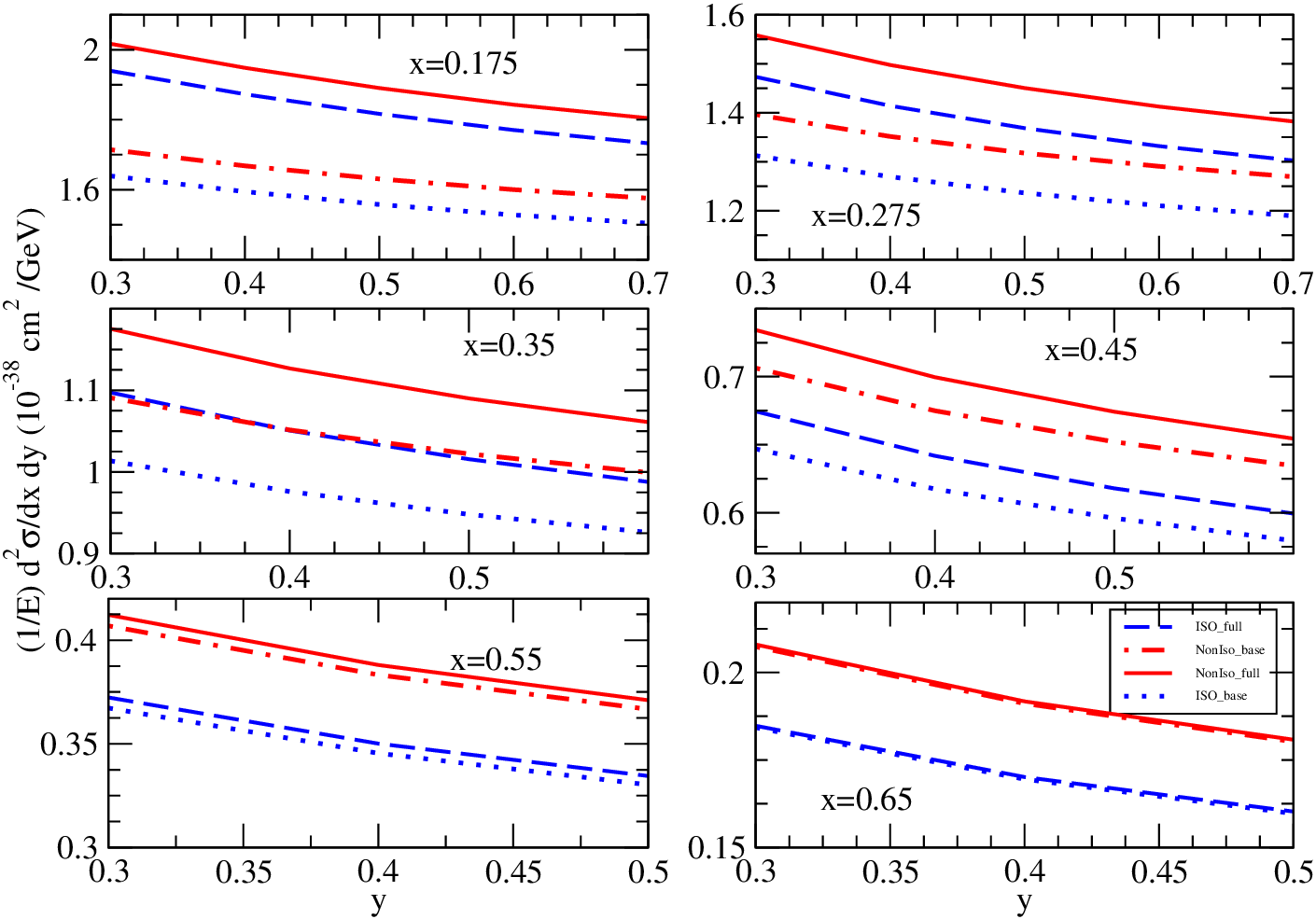}
\caption{(Color online) $\frac{1}{E}\frac{d^2\sigma}{dxdy}$ vs y at different x for $\nu_\mu$ induced reaction in $^{208}$Pb at $E_{\nu_\mu}=25$ GeV using CTEQ~\cite{cteq} PDF at LO. 
Isoscalar full model: dashed line; nonisoscalar one: solid line. 
Full model without pion, rho and shadowing with isoscalarity: dotted line; nonisoscalarity: dashed-dotted line.}
\label{iso_vs_noniso_nu}
\end{figure}
\begin{figure}
\includegraphics[scale=0.6]{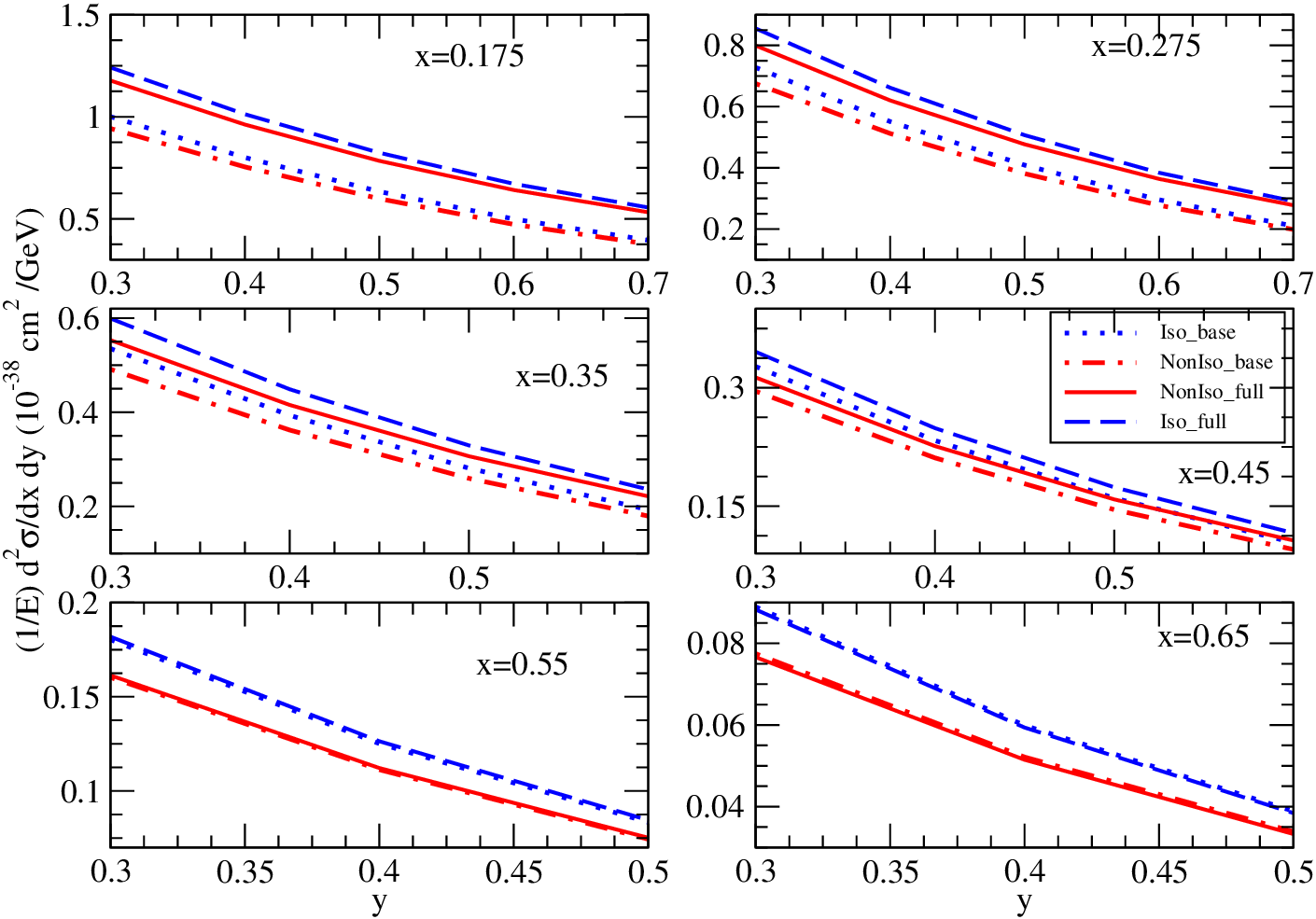}
\caption{ (Color online) $\frac{1}{E}\frac{d^2\sigma}{dxdy}$ vs y at different x for ${\bar\nu}_\mu$ induced reaction in $^{208}$Pb at $E_{{\bar\nu}_\mu}=25$ GeV using CTEQ~\cite{cteq} PDF at LO. 
Isoscalar full model: dashed line; nonisoscalar one: solid line. 
Full model without pion, rho and shadowing with isoscalarity: dotted line; nonisoscalarity: dashed-dotted line.}
\label{iso_vs_noniso_nubar}
\end{figure}
%\vspace{3mm}
\begin{figure}[h]
\includegraphics[width=12cm,height=12cm]{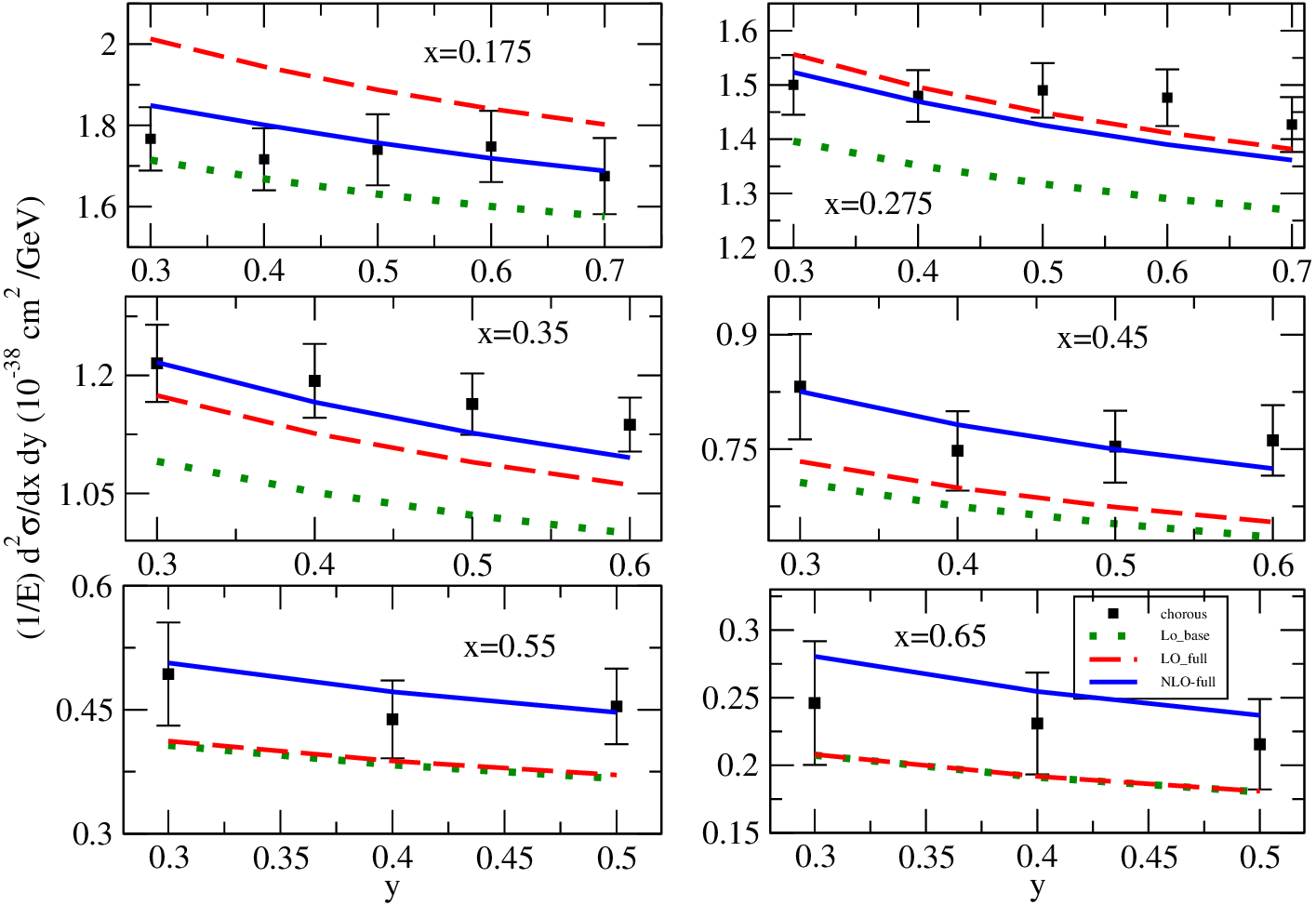}
\caption{ (Color online) $\frac{1}{E}\frac{d^2\sigma}{dxdy}$ vs y at different x for $\nu_\mu$($E_{\nu_\mu}=25$ GeV) induced reaction in $^{208}$Pb. Dotted line is the results using Eq.\ref{Diff_CS} with TMC. Solid (Dashed line) line 
is full calculation at NLO (LO). The experimental points are from CHORUS~\cite{chorus2}.}
\label{d2sigma_25gev_nu.eps}
\end{figure}
%\vspace{3mm}
\begin{figure}[h]
\includegraphics[width=12cm,height=12cm]{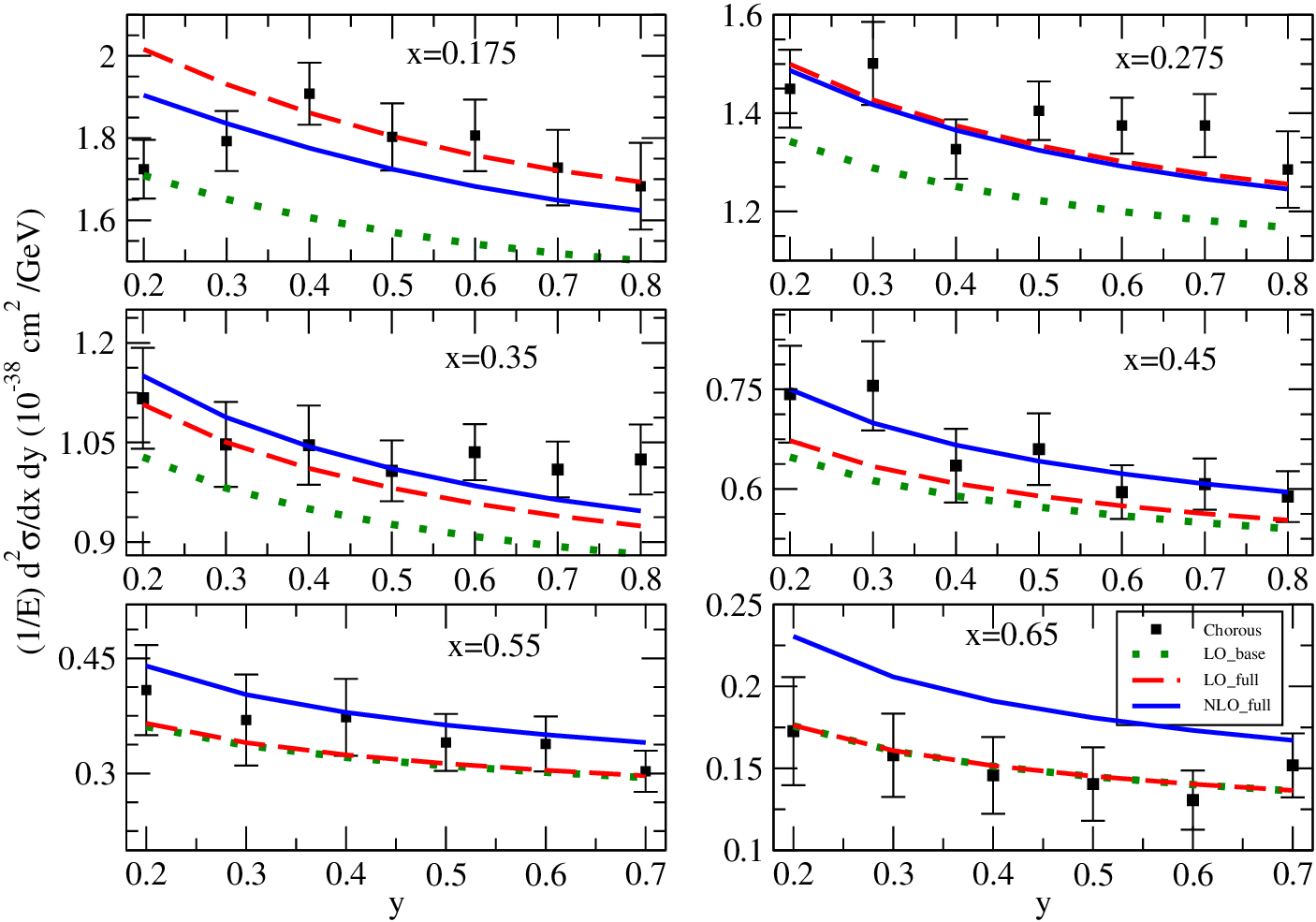}
\caption{(Color online) $\frac{1}{E}\frac{d^2\sigma}{dxdy}$ vs y at different x for $\nu_\mu$($E_{\nu_\mu}=70$ GeV) induced reaction in $^{208}$Pb. Lines and points have the same meaning as in Fig.\ref{d2sigma_25gev_nu.eps}.}
\label{d2sigma_70gev_nu.eps}
\end{figure}
%\vspace{3mm}
\begin{figure}[h]
\includegraphics[width=12cm,height=12cm]{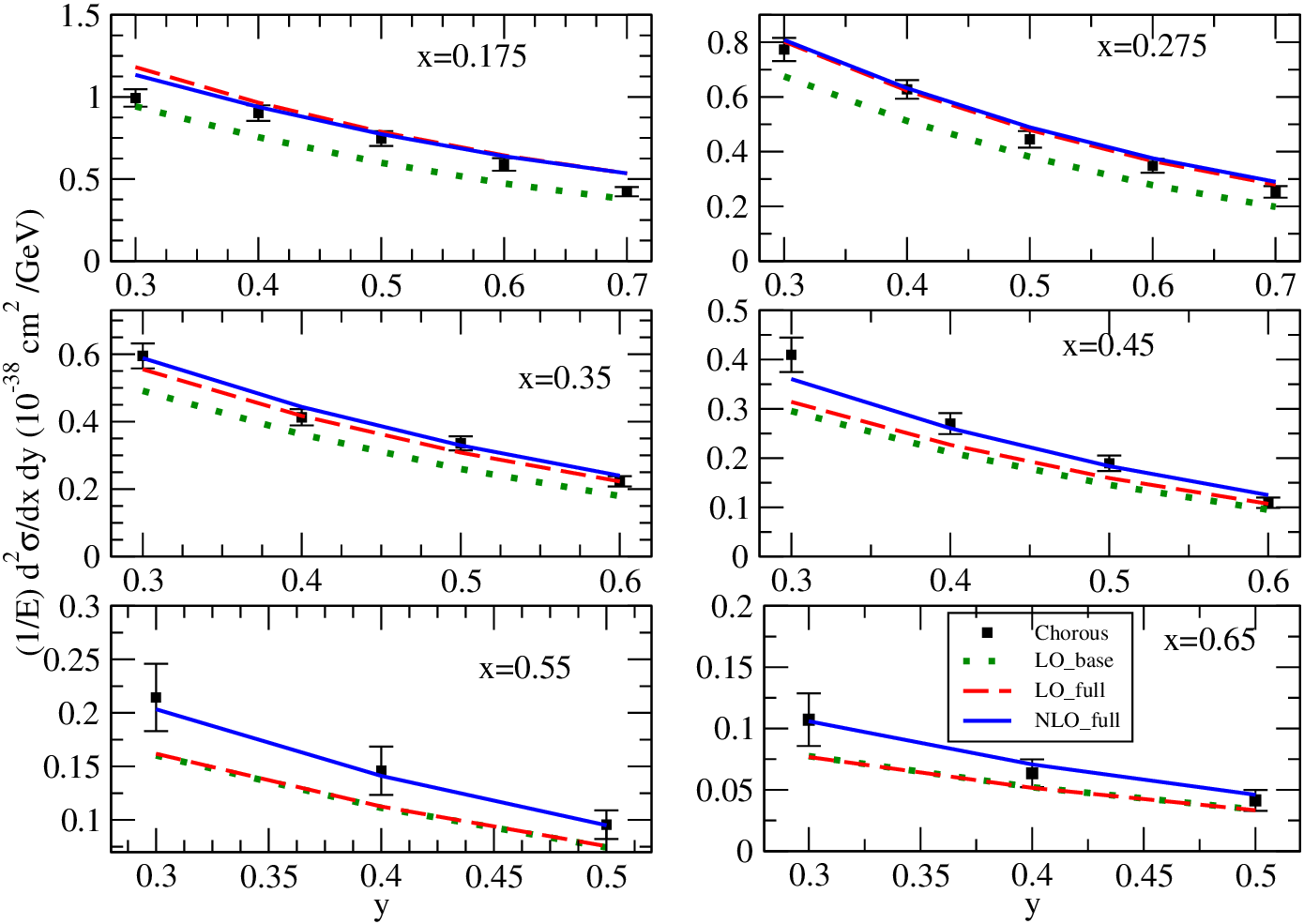}
\caption{(Color online) $\frac{1}{E}\frac{d^2\sigma}{dxdy}$ vs y at different x for $\bar\nu_\mu$($E_{\nu_\mu}=25$ GeV) induced reaction in $^{208}$Pb. Lines and points have the same meaning as in Fig.\ref{d2sigma_25gev_nu.eps}.}
\label{d2sigma_25gev_nubar.eps}
\end{figure}
\begin{figure}[h]
\includegraphics[width=12cm,height=12cm]{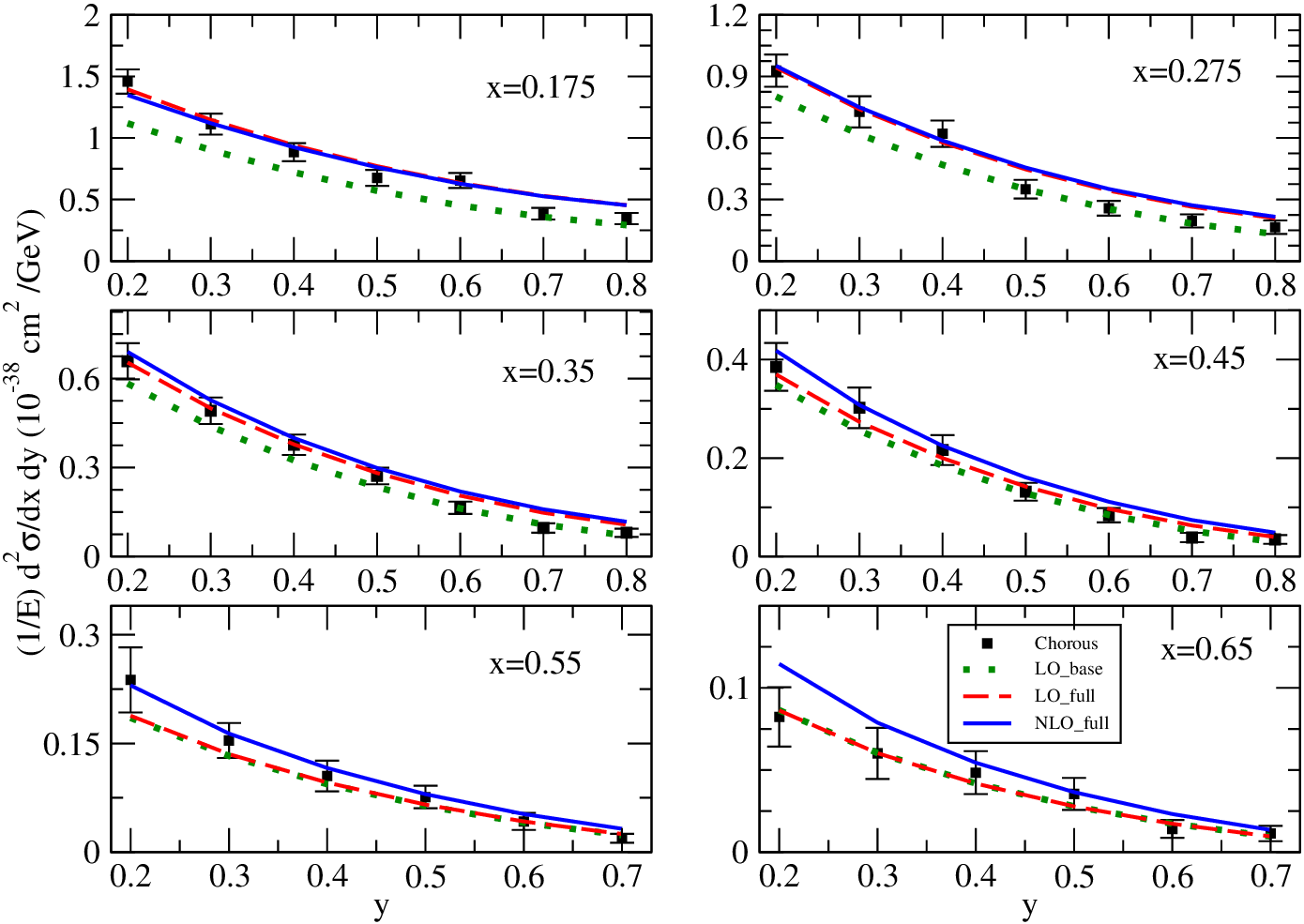}
\caption{(Color online) $\frac{1}{E}\frac{d^2\sigma}{dxdy}$ vs y at different x for $\bar\nu_\mu$($E_{\nu_\mu}=70$ GeV) induced reaction in $^{208}$Pb. Lines and points have the same meaning as in Fig.\ref{d2sigma_25gev_nu.eps}. }
\label{d2sigma_nubar_70gev.eps}
\end{figure}
Using Eqs.(\ref{f2Anuclei}) and (\ref{f3Anuclei}), we have calculated the $F^A_2$ and $F^A_3$ structure functions for the lead nucleus with target mass correction (TMC) \cite{schienbein} and 
CTEQ6.6 parton distribution functions (PDFs) at the Leading-Order (LO)~\cite{cteq}. We call this as our base result. Hereafter we include pion and rho cloud contributions in $F^A_2$ following the model of Ref. \cite{Marco} and the 
shadowing corrections in $F^A_2$ and $F^A_3$ \cite{Petti, Kulagin}, which we call as our full calculation (Total). These prescriptions have also been discussed in brief in our earlier works~\cite{Sajjad2,Sajjad1}. 
In Figs.(\ref{f2_ratio_lo.eps}) and (\ref{f3_ratio_lo.eps}), we have presented the results for the
 ratios $R_i(x,Q^2)$=$\frac{12F_{i}^{Pb}}{208F_i^C}$ and $\frac{56F_{i}^{Pb}}{208F_i^{Fe}}$ ($i=2,3$) at $Q^2=5$ GeV$^2$ and compared them with the results obtained from the 
phenomenological studies of Hirai et al.~\cite{Hirai} and Eskola et al.~\cite{Eskola}. The details of our calculations in $^{12}$C and $^{56}$Fe have been given in Ref.\cite{Sajjad2}. 
Here $^{12}$C is a pure isoscalar nucleus while $^{56}$Fe is a nonisoscalar one.
 The results for the ratios may be measured by the ongoing MINER$\nu$A experiment
 as they are planning to perform the cross section measurements in a wide range of neutrino energies using $^{12}$C, $^{56}$Fe and $^{208}$Pb nuclear targets. We find that the results for the 
ratios obtained by using our base calculation and full calculation are different at low x$<$0.5 which is due to the fact that this is the region where the mesons cloud contribution dominates while for x$>$0.5, the only nuclear effects are Fermi motion and 
binding energy corrections. Furthermore, we find that the nature of the ratios for $F_2$ and $F_3$ are not the same while in the phenomenological studies~\cite{Hirai, Eskola} 
they are almost the same in the present studied region of x. We also observe that the EMC effect is more prominent for ratios of structure functions between a heavy nucleus such as $^{208}$Pb 
and a much lighter one like $^{12}$C than for ratios between a heavy nucleus ($^{208}$Pb) 
and a medium-size one like $^{56}$Fe. This fact can be observed clearly in Figs. (\ref{f2_ratio_nlo.eps}) and (\ref{f3_ratio_nlo.eps}).

In Figs.(\ref{f2_ratio_nlo.eps}) and (\ref{f3_ratio_nlo.eps}) we have presented the results for $\frac{2F_{i}^{Pb}}{208F_i^D}$, $\frac{4F_{i}^{Pb}}{208F_i^{He}}$, $\frac{12F_{i}^{Pb}}{208F_i^C}$, $\frac{16F_{i}^{Pb}}{208F_i^O}$ and $\frac{56F_{i}^{Pb}}{208F_i^{Fe}}$ ($i=2,3$)
at $Q^2=5$ GeV$^2$ and $Q^2=50$ GeV$^2$ using CTEQ PDFs at NLO~\cite{cteq}. These results are presented for base as well as full calculations. 
The deuteron structure functions have been calculated using the same formulae as for $^{12}$C in Ref.\cite{Sajjad2}, but performing the convolution 
with the deuteron wave function squared instead of the spectral function. We have used the parametrization given in Ref.\cite{Lacombe:1981eg} for the deuteron wave function of the 
Paris N-N potential.

The results for $F^A_2$ and $F^A_3$ structure functions in the lead nucleus have been shown in Figs.(\ref{f2.eps}) and (\ref{f3.eps}) where we have also shown the CHORUS
 experimental data~\cite{chorus1, chorus2} for a wide range of x and $Q^2$. The effect of shadowing is
 about 5-7$\%$ at x=0.1, Q$^2$=1-5 GeV$^2$ and 1-2$\%$ at x=0.2, Q$^2$=1-5 GeV$^2$ which dies out with the increase in x and Q$^2$.
 In the case of $F^A_2$ there are pion and rho cloud contributions. The pion contribution is very dominant in comparison to the rho contribution. Pion contribution 
is significant in the region of $0.1~<~x~<~0.4$. Thus, we find that the shadowing correction seem to be negligible as compared to 
the other nuclear effects. It is the meson cloud contribution which is dominant at low and intermediate x for $F_2$. In these figures we also show the results 
of our full calculation at NLO. We can observe clearly that for medium Bjorken $x\simeq 0.6$ there is no difference between the base calculation and the full model at LO which is because in this $x$-region the mesons cloud contribution has a negligible effect on the structure functions.
 We find an overall better agreement between the NLO calculation and the experimental data. 

We must point out here that in the case of lead nucleus the correction due to nonisoscalarity is large and therefore the differential cross sections obtained by using the prescription given in our earlier paper~\cite{Sajjad2} for 
the isoscalar case is not valid here. To observe the effect of nonisoscalarity, we have calculated $\frac{1}{E}\frac{d^2\sigma}{dxdy}$ given in Eq.(\ref{Diff_CS}) using Eqs.(\ref{f2Anuclei}) and (\ref{f3Anuclei}) for $^{208}$Pb 
at $E_{\nu_\mu,{\bar\nu}_\mu}$=25 GeV and using Eqs.(13) and (14) of Ref.\cite{Sajjad2} treating lead to be isoscalar, and the results for $\nu_\mu$ and ${\bar\nu}_\mu$ induced processes are respectively shown in Figs.(\ref{iso_vs_noniso_nu})
and (\ref{iso_vs_noniso_nubar}) for the full model as well as with the base term only. The cross sections have been calculated using CTEQ~\cite{cteq} PDFs at LO. The difference is small at low x but increases
 with the increase in the values of x. For example, we find that the nonisoscalar correction to be around 4-5$\%$ at x=0.175 which becomes 10-12$\%$ at x=0.65.

In Figs.(\ref{d2sigma_25gev_nu.eps}) and (\ref{d2sigma_70gev_nu.eps}), we have shown the results for $\frac{1}{E}\frac{d^2\sigma}{dxdy}$ in $^{208}$Pb at $E_{\nu_\mu}$=25 GeV and 70 GeV respectively. 
The calculations for the the double differential cross sections have been performed for $Q^2~>~1$ GeV$^2$. These results are presented for our base and full calculations at LO and the full
 calculations at NLO. Similarly in Figs.(\ref{d2sigma_25gev_nubar.eps}) and (\ref{d2sigma_nubar_70gev.eps}), 
we have shown the results for $\frac{1}{E}\frac{d^2\sigma}{dxdy}$ induced by antineutrinos in $^{208}$Pb at $E_{\bar{\nu}_\mu}$=25 GeV and 70 GeV respectively. It is difficult to disentangle the
 main effects that make the curves being different because in the cross section both structure functions are mixed. However, we can say that for neutrinos there is a clear difference 
between LO and NLO calculation while for antineutrinos this difference seems to be smaller. We also find that the results improve 
(in comparison to the experimental data) when we add the contribution of mesonic degrees of freedom to the base calculation and even more when we perform the calculation at NLO.
 
\section{Conclusion}\label{Sec:Conclusion}
To conclude, in this work we have studied nuclear-medium effects in the structure functions $F^A_{2} (x, Q^2)$ and $F^A_{3} (x, Q^2)$ in lead nucleus using a many-body theory to describe 
the spectral function of the nucleon in the nuclear-medium for all $Q^2$. The local density approximation has been used to apply the results for the finite nuclei. The use of the spectral 
function is to incorporate Fermi motion and binding effects. We have used CTEQ~\cite{cteq} PDFs in the numerical evaluation. Target mass correction has been considered.
We have taken the effects of mesonic degrees of freedom, shadowing, anti-shadowing in the calculation of $F_2^A$ and shadowing and anti-shadowing effects in the calculation of $F^A_{3}$. 
We have found that the mesonic cloud (mainly pion) gives an important contribution to the cross section. These numerical
results have been compared with the experimental observations of CHORUS~\cite{chorus2} Collaboration. Using these structure functions we obtained differential scattering cross section 
for lead and compared the results with the CHORUS~\cite{chorus2} observation. The results for the structure functions in carbon and iron~\cite{Sajjad2}, helium and oxygen have also been used to study the ratio
 $\frac{2F_{i}^{Pb}}{208F_i^D}$, $\frac{4F_{i}^{Pb}}{208F_i^{He}}$, $\frac{12F_{i}^{Pb}}{208F_i^C}$, $\frac{16F_{i}^{Pb}}{208F_i^O}$ and $\frac{56F_{i}^{Pb}}{208F_i^{Fe}}$ (i=2,3),  where for deuteron we have used the same formulae as for $^{12}$C in Ref.\cite{Sajjad2}, 
but performing the convolution with the deuteron wave function squared instead of the spectral function. These results have been compared with some of the phenomenological studies by Hirai et al~\cite{Hirai} and 
Eskola et al.~\cite{Eskola}. We find that the effect of the nuclear-medium is also quite important even for the deep inelastic scattering. Furthermore, we observe that the ratio of the 
structure functions for the different nuclei are not the same. MINER$\nu$A~\cite{minerva} would be able to measure these ratios as they are using deuterium, helium, carbon, iron and 
lead targets and our results may be quite useful in their analysis.
\begin{acknowledgments}
The authors thank M. J. Vicente Vacas, University of Valencia, Spain for many useful discussions and encouragement throughout this work. This research was supported by DGI
and FEDER funds, under contracts FIS2008-01143/FIS, FIS2006-03438, and the Spanish Consolider-Ingenio 2010
Program CPAN (CSD2007-00042), 
by Generalitat Valenciana contract PROMETEO/2009/0090 and by the EU
HadronPhysics2 project, Grant Agreement No. 227431. M. S. A. wishes to acknowledge the financial support from the University of Valencia and the Aligarh Muslim University under the academic
 exchange program and also to the DST, Government of India for financial support under the grant SR/S2/HEP-0001/2008. H. H. acknowledges the Maulana Azad National Program. 
\end{acknowledgments}

\end{document}